\documentclass{sig-alternate}
\pdfoutput=1
\usepackage{etoolbox}
\makeatletter
\patchcmd{\maketitle}{\@copyrightspace}{}{}{}
\makeatother
\usepackage{url}
\usepackage{flushend}
\usepackage[T1]{fontenc}
\usepackage{array}
\newcolumntype{P}[1]{>{\raggedright\arraybackslash}p{#1}}
\usepackage{verbatim}
\usepackage[hidelinks]{hyperref}
\usepackage{color}
\usepackage{listings}
\usepackage[hang,flushmargin]{footmisc} 
\renewcommand{\footnoterule}{\vspace*{-3pt}%
    \hspace{-0.04\columnwidth}\rule{\columnwidth}{0.2pt}\vspace*{2.6pt}}

\begin{document}
\renewcommand\footnotemark{}
\title{A Taxonomy for Attack Patterns on Information Flows in Component-Based Operating Systems}

%
%
%
%
%

\numberofauthors{2} 
%
\author{
\alignauthor
Michael Hanspach\\
       \affaddr{Fraunhofer FKIE, Wachtberg, Germany}\\
       \email{michael.hanspach@fkie.fraunhofer.de}
\alignauthor
J{\"o}rg Keller\\
       \affaddr{FernUniversit{\"a}t in Hagen, Germany}\\
       \email{joerg.keller@fernuni-hagen.de}
}

\maketitle

\begin{abstract}
We present a taxonomy and an algebra for attack patterns on component-based operating systems.
In a multilevel security scenario, where isolation of partitions containing data at different security classifications is the primary security goal and security breaches are mainly defined as undesired disclosure or modification of classified data, strict control of information flows is the ultimate goal.
In order to prevent undesired information flows, we provide a classification of information flow types in a component-based operating system and, by this, possible patterns to attack the system.
The systematic consideration of informations flows reveals a specific type of operating system covert channel, the covert physical channel, which connects two former isolated partitions by emitting physical signals into the computer's environment and receiving them at another interface.
\end{abstract}




\section{Introduction}
Operating systems for trustworthy computing systems often consist of a verifiable (micro) kernel and individual service components.
Jaeger describes this type of operating system as a component-based operating system~\cite{Jaeger:1998:SAC:319195.319229}.

Rushby's separation kernel provides isolated regimes for components, with the goal ``to create an environment which is indistinguishable from
that provided by a physically distributed system''~\cite{Rushby:1981:DVS:800216.806586}.
The MILS architectural approach places ``traditional kernel-level security functionalities into
external modular components that are small enough for rigorous
security evaluation using formal methods'' as described by Robinson et al.~\cite{Robinson:2007:IMC:1314466.1314474}.

In a multilevel security scenario, where different partitions have different levels of classification and must be isolated to prevent information leakage, attacks on component-based operating systems might be successful that are based on exploiting shared resources, on colluding components or on exploiting the physical characteristics of the computing system.
In this paper, we aim to identify and classify the attack patterns on component-based operating systems by analyzing all information flows in the model of a component-based operating system.
An algebra for information flows is presented that could easily be adapted for the analysis of extended models of operating systems.
As part of our analysis, the security goals of confidentiality and integrity are targeted, while other security goals such as availability are not included.

Our contribution includes a taxonomy of attack patterns on information flows in component-based operating systems based on the performed analysis.
As a further result of the analysis, a specific type of operating system covert channel, namely the covert physical channel, is identified, which has not been extensively discussed to our knowledge and might demand increased awareness in the future.

The remainder of this paper is structured as follows.
In Section \ref{sec_ap2}, we present the basic scenario used in our work.
In Section \ref{sec_ap3}, we identify all legitimate and illegitimate information flows possible in the basic scenario. 
In Section \ref{sec_ap4}, we map the identified illegitimate information flows to attack patterns, which represent methods for establishing and exploiting illegitimate information flows.
In Section \ref{sec_ap_rl}, we describe how our work is connected to related work.
In Section \ref{sec_ap6}, we draw our conclusions.

\section{Basic Scenario}\label{sec_ap2}

\begin{figure}[ht]
  \begin{center}
  \includegraphics[width=0.3\textwidth]{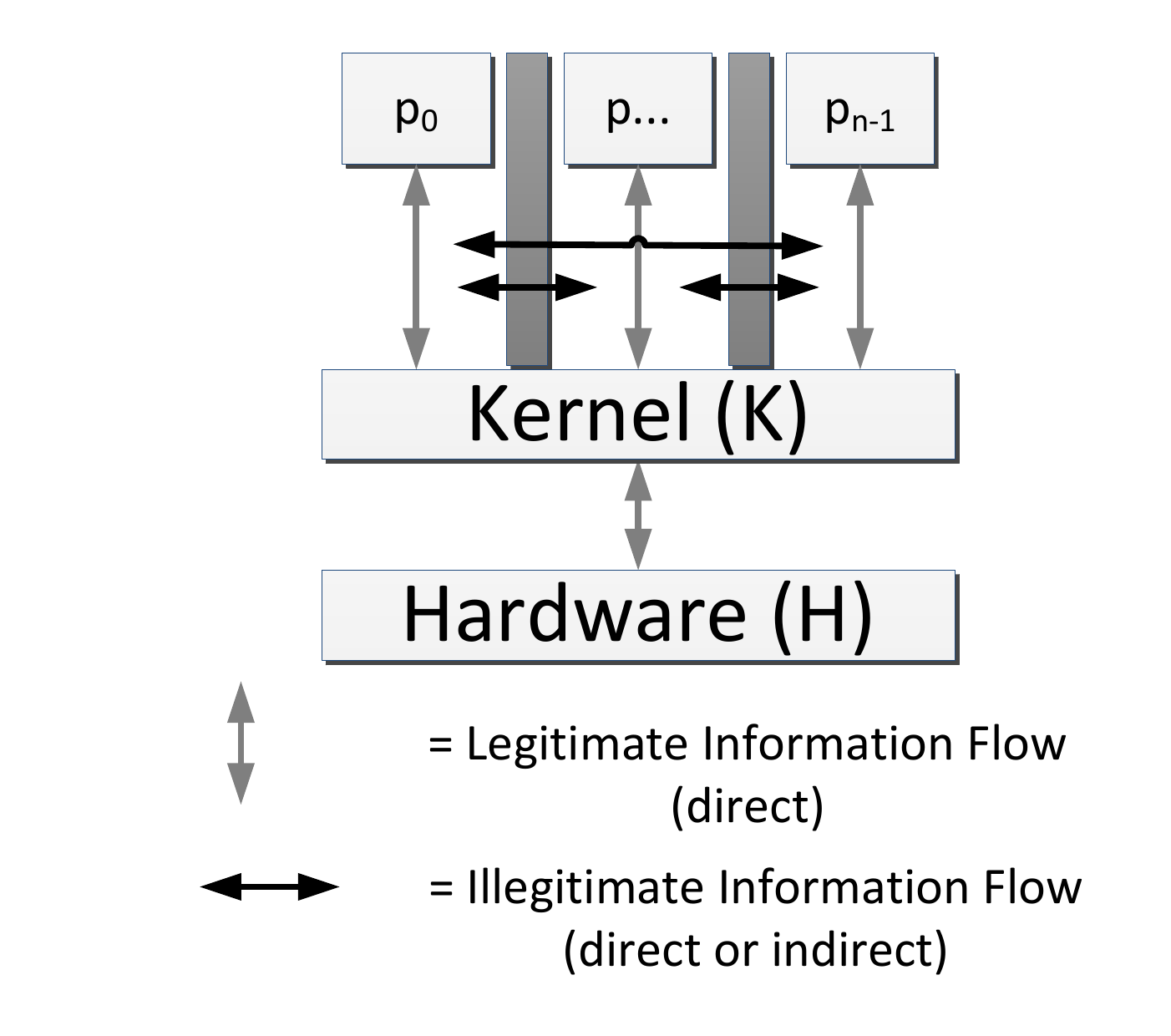}
  \end{center}
  \caption{Scenario for attacks on a component-based operating system}\label{scenario}
\end{figure}

The basic scenario used for the attack patterns described in this paper is shown in Fig.~\ref{scenario}.

We assume to have $n \geq 2$ isolated application partitions running on top of the (micro) kernel using shared hardware resources.
These partitions could contain virtual machines or individual applications that are directly accessible to a user of the computing system.

Micro kernels providing isolation mechanisms for multilevel security can roughly be differentiated into \emph{sharing hypervisors} and \emph{pure isolation hypervisors}, where sharing hypervisors do allow communication between application partitions under certain circumstances and pure isolation hypervisors do not allow any communication between these partitions at all \cite{Karger05multi-levelsecurity}.
In this paper, only attack patterns on pure isolation hypervisors are considered, leaving attack patterns on the sharing mechanisms between partitions as a subject for further research. 

Although a multilevel security scenario is technically not necessary to describe attack patterns on the isolation mechanisms of a component-based operating system, it can serve the purpose to back up the theory with possible applications and examples.

The micro kernel is implementing the reference monitor concept, which states that ``the reference validation mechanism must always be invoked'' \cite{anderson:rm}.
In other words: any information flow has to be governed by the reference monitor respectively the micro kernel, in this scenario. 
Still, micro kernels are usually unable to really govern \emph{any} information flow, as there may be undesired channels left in the system. 
Lampson \cite{Lampson:1973:NCP:362375.362389} differentiated between three basic types of channels: legitimate channels, storage channels and covert channels.
As only the legitimate channels are governed by the kernel, storage channels as well as covert channels can be used to establish additional undesired information flows.
We discuss each of these channel types in the course of this paper and analyze their very different subcategories.

We will start with a study of information flows in the basic scenario.
\section{Information Flows}\label{sec_ap3}
\subsection{Basic Information Flows}\label{basic_flows}

\begin{figure}[ht]
  \begin{center}
  \includegraphics[width=0.5\textwidth]{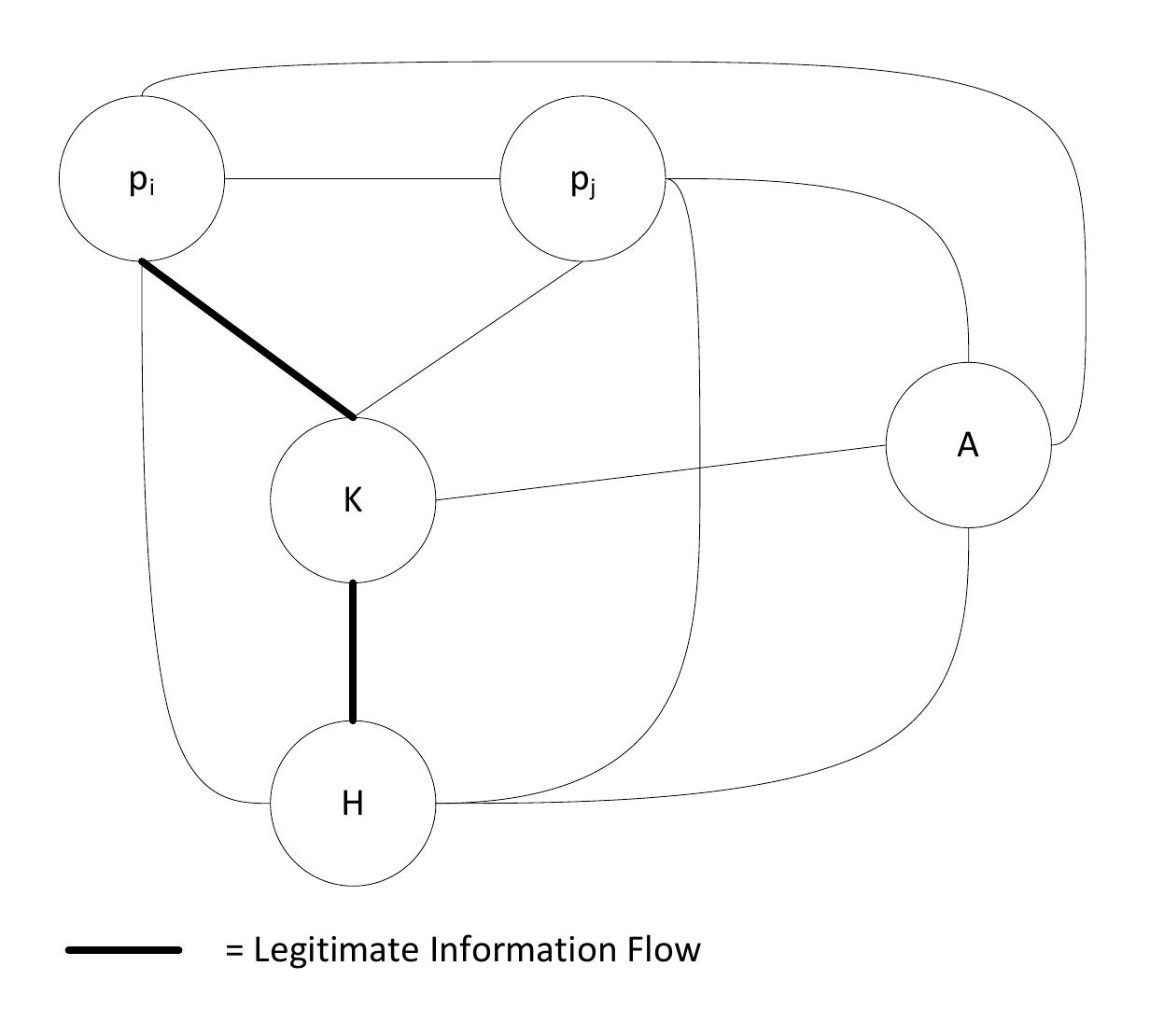}
  \end{center}
  \caption{Graph of information flows in a component-based operating system\label{graph}}
\end{figure}

Let $P=\{p_0,\ldots,p_{n-1}\}$ represent the set of isolated application partitions running on top of the kernel.
Let $K$ represent the operating system kernel and
let $H$ represent the underlying hardware.
Finally, let $A$ be the attacker, seeking to read or modify data without the proper authorization.
With these variables, we can construct a complete graph (Fig.~\ref{graph}) representing all possible information flows in our scenario as edges or sequences of incident edges in the graph.

To define the legitimate information flows from the set of all possible information flows, we start a walk at $p_i$.
As any information flow between $p_i$ and $p_j (i \neq j)$ is declared illegitimate in a pure isolation hypervisor, we cannot walk to $p_j$.
Furthermore, $p_i$ is not allowed to communicate with the attacker and direct access to the hardware is illegitimate, as any information flow has to be governed by the always-invoked reference monitor, which is $K$ in this scenario. 
Therefore, $p_i$ is only able to directly communicate to $K$, and $H$ may only be accessed over $K$.
After walking to $K$, we can walk further to $H$ or we could walk back to $p_i$, as any information flow can be implemented bidirectionally.
It should be noted that walking from $K$ to $p_j$ is illegitimate when starting at $p_i$, although information flows between $K$ and $p_j$ are generally legitimate.

To demonstrate the identified information flows, we define $\Longleftrightarrow$ as a bidirectional direct information flow between two vertices.
Likewise, $\Longrightarrow$ is defined as a unidirectional direct information flow from left to right between two vertices.
As explained above, the only legitimate information flows in our scenario (starting from $p_i$) are:

\begin{enumerate}
\item $p_i \Longleftrightarrow K$

\item $K  \Longleftrightarrow H$
\end{enumerate}

While we are walking forth from a specific starting position (e.g. $p_i$), we can concatenate these information flows.
Therefore, the following information flow is also legitimate:

\begin{enumerate}
\setcounter{enumi}{2}
\item $p_i \Longleftrightarrow K \Longleftrightarrow H$
\end{enumerate}

We can identify any illegitimate information flow in the model with a recursive walk algorithm as shown in Listing~\ref{walk_algo} (pseudocode).

\begin{figure}
\begin{lstlisting}[caption=Recursive walk algorithm,label=walk_algo,language=perl]
 function walk-to-neighbors (start, walk)

	for all neighbors of start do

			local-walk := concatenate(walk, neighbor)
			walk-legitimate := test-walk(start, neighbor)

			if (walk-legitimate = true)
			then
				walk-to-neighbors(neighbor, local-walk)
			else
				store local-walk
			end-if

	end-for

 end-function
\end{lstlisting}
\end{figure}

In Listing 1, line 1, the function and its arguments are declared where \emph{start} refers to the position in the graph where a local walk is started and \emph{walk} refers to the complete walk that has been performed before calling the function.
In the loop between lines 3 and 15, any possible walk to a neighbor is tested (line 6) if it conforms to the legitimate walks (i.e. information flows).
If a walk to a neighbor is legitimate, the function is recursively called again (line 10), while the id of the neighbor and the previously performed walk (line 5) are given as the new arguments.
If a walk is found to be illegitimate (line 12), the illegitimate information flow is stored and the algorithm continues with the next neighbor that is addressed in the loop.

By calling the algorithm with $p_i$ as the start position and an empty walk, we can catch any shortest form of an illegitimate information flow from $p_i$ to other vertices.
As the algorithm stops as soon as an information flow turns out as illegitimate, one could produce even longer information flows by walking forth from the last vertex, but the resulting information flows would just be extended variants of the hereby identified information flows.
So, if any of these identified illegitimate information flows is part of an even longer information flow, this information flow is also considered an illegitimate information flow.
Therefore, for demonstrating a channel between $p_i$ and $p_j$, one can always append $\Longleftrightarrow p_j$ to the illegitimate information flow, if not already present.
Likewise, for demonstrating an attack, $\Longleftrightarrow A$ can be appended, and to show a unidirectional information flow, $\Longleftrightarrow$ might be replaced by $\Longrightarrow$ in every information flow.
We verified the resulting illegitimate information flows by implementing the algorithm in the programming language Perl.
The resulting illegitimate information flows for the basic scenario are listed below.

\begin{enumerate}
\item $p_i \Longleftrightarrow A$

\item $p_i \Longleftrightarrow p_j$

\item $p_i \Longleftrightarrow H$

\item $p_i \Longleftrightarrow K \Longleftrightarrow A$

\item $p_i \Longleftrightarrow K \Longleftrightarrow p_j$

\item $p_i \Longleftrightarrow K \Longleftrightarrow H \Longleftrightarrow p_j$

\item $p_i \Longleftrightarrow K \Longleftrightarrow H \Longleftrightarrow A$
\end{enumerate}

As discussed in the last section, a kernel implementing the reference monitor concept must always be invoked.
Therefore, any communication in the form of $p_i \Longleftrightarrow p_j$ should be impossible by definition, as the kernel would not be invoked in this information flow.
Still, attacks using the $p_i \Longleftrightarrow p_j$ approach have to be considered, as there may be implementation flaws in the kernel or in the hardware or the system may be misconfigured by a system administrator.
These types of attack can only be consequently circumvented by formal verification of hardware, software and the implemented mandatory access control (MAC) policy, describing a policy that ``enforces
systemwide security invariants regardless of user preference''~\cite{watson_mac} (e.g. non-interference as described by Goguen and Meseguer~\cite{goguen}) in contrast to a discretionary access control (DAC) policy.

It should also be obvious that the hardware is involved in any physical information flow within the operating system.
In contrast to this, we are only specifying logical information flows that only include the components which are actually behaving contrary to the defined policy.
So, as in $p_i \Longleftrightarrow K \Longleftrightarrow p_j$, $K$ may be malfunctioning, while $H$ is running just as expected and, therefore, $K$ is defined as the cause of the information leak.
Likewise, in $p_i \Longleftrightarrow K \Longleftrightarrow H \Longleftrightarrow A$, $H$ is defined as the cause of the information leak, while $K$ may be running perfectly fine.

\subsection{Using operating system guards}
Even in a pure isolation hypervisor, there must be some kind of communication between the partitions of system components and the application partitions.
Service components provide services, maybe a file system implementation or a network service, which are not integrated into the (micro) kernel.
These service components can be categorized into two different types: untrusted (i.e. non-trustworthy) components and trusted (i.e. trustworthy) components.\hfill\newline
Trustworthy components perform security-critical tasks in a component-based operating system, for instance, multiplexing of shared resources, cryptographic operations and integrity checks.
Trustworthy components have to be small enough in code size to be subject to evaluation or formal verification.
We shall refer to trustworthy components that provide services to other components as \emph{guards}, with the set of guards defined as: $G=\{g_0,\ldots,g_{m-1}\}$.
The set of non-guard-components, defined as $C=\{c_0,\ldots,c_{q-1}\}$, may be guarded (i.e protected) by those guards.
An untrusted component $c \in C$ could, for instance, be a driver component, reutilizing existing (untrusted) code to access a shared hardware device.

\begin{figure}[!ht]
\begin{center}
\includegraphics[width=0.5\textwidth]{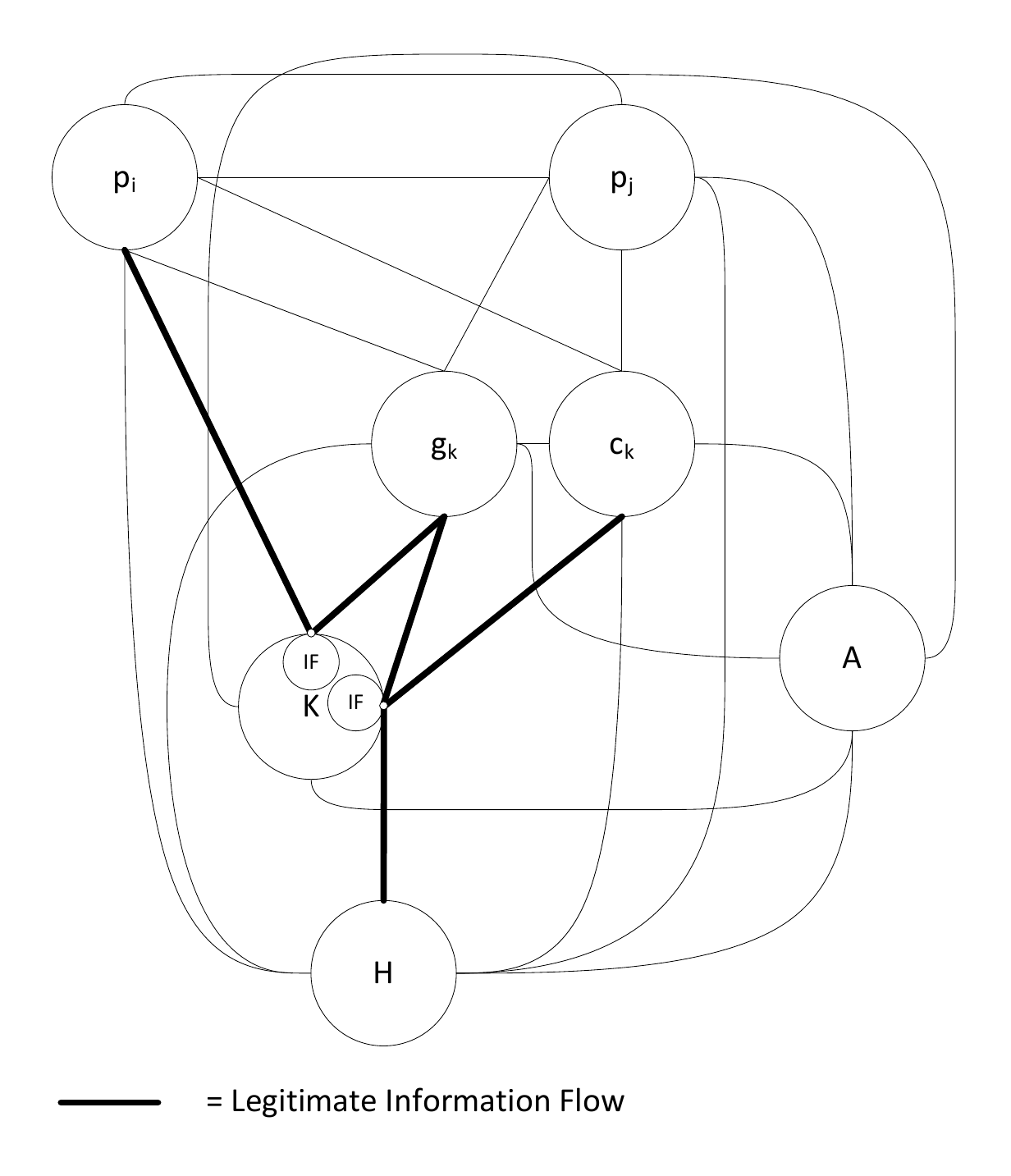}
\end{center}
\caption{Using operating system guards}\label{graph2}
\end{figure}

Extending the scenario to guards and guarded components (i.e. untrusted components that are protected by a guard) leads to a more complex version of the graph (Fig.~\ref{graph2}).
The information flows in the extended scenario can be described as follows:

A shared hardware resource $H$ may only be accessed via a guarded component $c_k$ or via a guard $g_k$.
The guard $g_k$ provides isolated resource access to $p_i$.
All of these information flows have to be governed by $K$.

This description leads us to the following legitimate information flows:

\begin{enumerate}
\item $c_k \Longleftrightarrow K \Longleftrightarrow H$

\item $g_k \Longleftrightarrow K \Longleftrightarrow c_k$

\item $p_i \Longleftrightarrow K \Longleftrightarrow g_k$

\item $g_k \Longleftrightarrow K \Longleftrightarrow H$
\end{enumerate}

As described in Section~\ref{basic_flows}, we are able to concatenate the legitimate information flows to even longer legitimate information flows.
The resulting legitimate information flows of components and guards can be formalized as:

\begin{enumerate}
\setcounter{enumi}{4}
\item $p_i \Longleftrightarrow  K \Longleftrightarrow g_k \Longleftrightarrow K \Longleftrightarrow c_k \Longleftrightarrow K \Longleftrightarrow H$
\item $p_i \Longleftrightarrow  K \Longleftrightarrow g_k \Longleftrightarrow K \Longleftrightarrow H$
\end{enumerate}

For better visualization of the legitimate information flow, we introduce interfaces (IF) to $K$ in Fig.~\ref{graph2}.
Only those vertices connected to the same interface of $K$ may actually exchange information over the kernel.
So, while information flows accessing the shared hardware resources via a guard component and thereby isolating the application partitions are legitimate, any other information flows are illegitimate in this example.

Illegitimate information flows for an extended scenario (using operating system guards) can be retrieved with the same algorithm as used in Section~\ref{basic_flows}.
We will have a further look at one of these illegitimate information flows that describes a new type of attack pattern (as listed in Section~\ref{sec_ap4}).
\begin{enumerate}
\item $p_i \Longleftrightarrow K \Longleftrightarrow c_k \Longleftrightarrow K \Longleftrightarrow p_j$
\end{enumerate}
Due to shared access to the untrusted component $c_k$, $p_i$ and $p_j$ may be able to exchange information among each other and, therefore, violate the MAC policy.

Now, with the illegitimate information flows identified, we are able to identify the very different attack patterns to a component-based operating system.
The next section will describe these attack patterns and the process of identifying them.

\section{Attack Patterns}\label{sec_ap4}

We can map the information flows identified in the previous section to attack patterns.
Attack patterns are defined as methods for establishing or exploiting illegitimate information flows.

The relation between information flows and attack patterns can be described as a many-to-many relation:
Multiple attack patterns can be mapped to a single information flow, as there may be very different methods for establishing or exploiting an illegitimate information flow.
Likewise, multiple information flows can be mapped to a single attack pattern, as one method for exploiting or establishing illegitimate information flows could be used for multiple information flows.
A mapping between information flows and basic attack patterns is established in Tab.~\ref{table_flows}. 

\begin{table*}[hp]
\centering
\begin{tabular}{ | c | p{6cm} | l | }
\hline                        
\textbf{No.} & \textbf{Information Flow} & \textbf{Basic Attack Patterns}\\
\hline                        
1 & $p_i \Longleftrightarrow A$ & Use insider attack \\
\hline       
2 & $p_i \Longleftrightarrow p_j$ & Exploit flaws in the system policy \\
\hline                        
3 & $p_i \Longleftrightarrow H$ & Use physical attacks, Exploit covert channels, Exploit storage channels \\
\hline                        
4 & $p_i \Longleftrightarrow K \Longleftrightarrow A$ & Exploit flaws in the system policy \\
\hline                        
5 & $p_i \Longleftrightarrow K \Longleftrightarrow p_j$ & Exploit covert channels, Exploit flaws in the system policy \\
\hline                        
6 & $p_i \Longleftrightarrow K \Longleftrightarrow H \Longleftrightarrow p_j$ & Use physical attacks, Exploit covert channels \\
\hline                        
7 & $p_i \Longleftrightarrow K \Longleftrightarrow H \Longleftrightarrow A$ & Use physical attacks \\
\hline  
8 & $p_i \Longleftrightarrow K \Longleftrightarrow c_k \Longleftrightarrow K \Longleftrightarrow p_j$ & Exploit flaws in the system policy \\
\hline  
\end{tabular}
\caption{Illegitimate information flows mapped to basic attack patterns\label{table_flows}}
\end{table*}

Additionally, it should be noted that the attack pattern of exploiting implementation flaws is associated with any of these information flows, as flaws might always be present in hardware or software that has not been subject to formal verification.
The basic attack patterns presented can be further differentiated along the specific methods used in an attack in order to generate more precise (extended) attack patterns (see Tab.~\ref{ext_patt}).

\begin{table*}[hp]
\centering
\begin{tabular}{ | c | p{6cm} | p{10cm} | }
\hline                        
\textbf{No.} & \textbf{Basic Attack Pattern} & \textbf{Extended Attack Patterns} \\
\hline                        
1 & Use insider attack & N/A \\
\hline       
2 & Use physical attacks & Exploit physical access, Exploit side channels \\
\hline                        
3 & Exploit implementation flaws & N/A \\
\hline                        
4 & Exploit covert channels & Exploit covert storage channels, Exploit covert timing channels, Exploit covert physical channels \\
\hline  
5 & Exploit storage channels & N/A \\
\hline                        
6 & Exploit flaws in the system policy & Exploit corrupt policy channels, Exploit illegitimate access to components \\
\hline  
\end{tabular}
\caption{Basic attack patterns mapped to extended attack patterns\label{ext_patt}}
\end{table*}

The difference between basic and extended attack patterns is that extended attack patterns utilize the same information flows as their associated basic attack pattern, but they each describe a distinct method to establish the information flow.
It should further be noted that in order to describe side channels and covert physical channels, $\Longleftrightarrow$ should be replaced by $\Longrightarrow$ as the information flow is only unidirectional in most cases.
Finally, the identified attack patterns are described in Tab.~\ref{ext_patt2}.
Where extended attack patterns have been identified, only the extended attack patterns are listed.
\begin{table*}[hp]
\centering
\begin{tabular}{ | c | P{4cm} | P{8cm} | P{3,5cm} | }
\hline                        
\textbf{No.} & \textbf{Attack Pattern} & \textbf{Description} & \textbf{References} \\
\hline                        
1 & Use insider attack & $A$ is granted direct access to extract information from $p_i$. & Baracaldo and Joshi~\cite{Baracaldo:2013:BAU:2462410.2462411}, Liu et al.~\cite{Liu:2008:DSD:1413140.1413159}, Yu and Chiueh~\cite{Yu:2004:DFS:1029146.1029154}. \\
\hline       
2 & Exploit physical access & $A$ has physical access to the computing system and can directly extract information, which has been stored at $H$, from $p_i$. & Halderman et al.~\cite{Halderman:2009:LWR:1506409.1506429}. \\
\hline  
3 & Exploit side channel & $A$ uses side channel attacks to extract information from $p_i$ over $H$. & van Eck~\cite{vanEck:1985:ERV:7307.7308}, D{\"u}rmuth~\cite{Duermuth09}, Backes et al.~\cite{Backes:2010:ASA:1929820.1929847}, LeMay and Tan~\cite{lemay}, Shamir and Tromer~\cite{cryptana}, Halevi and Saxena~\cite{Halevi:2012:CLK:2414456.2414509}. \\
\hline                        
4 & Exploit implementation flaws & $A$ uses implementation flaws of $p_i$, $K$ or $H$ to extract information from $p_i$ & Klein et al.~\cite{Klein:2009:SFV:1629575.1629596}, Boettcher et al.~\cite{4702758}, Robinson et al.~\cite{Robinson:2007:IMC:1314466.1314474} \\
\hline                        
5 & Exploit covert storage channel & A covert channel is established between $p_i$ and $p_j$ by storing a hidden message within shared resources. & Lampson \cite{Lampson:1973:NCP:362375.362389}, National Computer Security Center~\cite{rainbow}. \\
\hline  
6 & Exploit covert timing channel & A covert channel is established between $p_i$ and $p_j$ by encoding a message via the timing behavior of shared resources. & Lampson \cite{Lampson:1973:NCP:362375.362389}, National Computer Security Center~\cite{rainbow}. \\
\hline  
7 & Exploit covert physical channel & A covert physical channel is established between $p_i$ and $p_j$ by encoding a message, sending it out into the physical environment via $H$ and receiving it at a different interface of $H$. & N/A. \\
\hline  
8 & Exploit storage channels & Information can be exchanged between $p_i$ and $p_j$ via shared storage resources. & Lampson \cite{Lampson:1973:NCP:362375.362389} \\
\hline                        
9 & Exploit corrupt policy channel & Information can be exchanged directly between $p_i$ and $p_j$ due to a flawed MAC policy. & Agreiter~\cite{Agreiter:2008:MCO:1370175.1370221}, Zhai et al.~\cite{Zhai:2009:DIT:1655925.1656007}. \\
\hline  
10 & Exploit illegitimate access to components & Information can be exchanged between $p_i$ and $p_j$ by exploiting illegitimate access to a component. &  N/A \\
\hline  
\end{tabular}
\caption{Description and references regarding to attack patterns\label{ext_patt2}}
\end{table*}
A discussion of the provided references is included in Section~\ref{sec_ap_rl} (related work).

We also introduce the concept of a \emph{covert physical channel}, which uses physical signals that were not meant for communication in the first place, to transmit covert messages between isolated application partitions.
Preliminary work on physical communication has been performed by Madhavapeddy et al.~\cite{Madhavapeddy:2005:ANF:1083818.1083942}, who present a study on audio networking, by Loughry and Umphress~\cite{Loughry:2002:ILO:545186.545189}, who describe information leakage from optical emanations, by Hasan et al.~\cite{Hasan:2013:SCH:2484313.2484373}, who study different physical means for command-and-control communication between mobile devices and by Raguram et al.~\cite{Raguram:2011:IAR:2046707.2046769}, who present a setup where typed keystrokes are extracted by optical reflections of the environment.
In contrast to these authors, we are not just looking at physical means for communication or information extraction, but we identify a covert channel between two application partitions that is established by sending physical (e.g. optical or acoustical) signals into the computer's environment and receiving them at another interface of the computing system.

Karger and Wray~\cite{kargerwray} present a covert storage channel based on disk arm optimization.
This covert channel would not qualify as a covert physical channel, because the physical environment (shared by different partitions) would not be used for the covert channel and the covert channel only takes place at the hardware level.
This difference can be made visible by introducing a variable $E$ (describing the shared physical environment) to the model. 

Where the disk arm channel might be described by:\newline
$p_i \Longleftrightarrow K \Longleftrightarrow H \Longleftrightarrow K \Longleftrightarrow p_j$,\newline
The covert physical channel would be better described by:\newline
$p_i \Longleftrightarrow K \Longleftrightarrow H \Longleftrightarrow E \Longleftrightarrow H \Longleftrightarrow K \Longleftrightarrow p_j$.

The variable $E$ might be used in future analyses to identify a covert channel as a type of covert physical channel.

Murdoch~\cite{Murdoch:2006:HRH:1180405.1180410} presents a covert channel based on clock skew manipulations as a result of different heat output levels.
This type of covert channel would also not qualify as a covert physical channel according to our terminology, because the signal is not directly transferred between sender and receiver over a shared physical environment, but indirectly by clock skew manipulations and analysis.

\begin{figure}[!ht]
\includegraphics[width=0.5\textwidth]{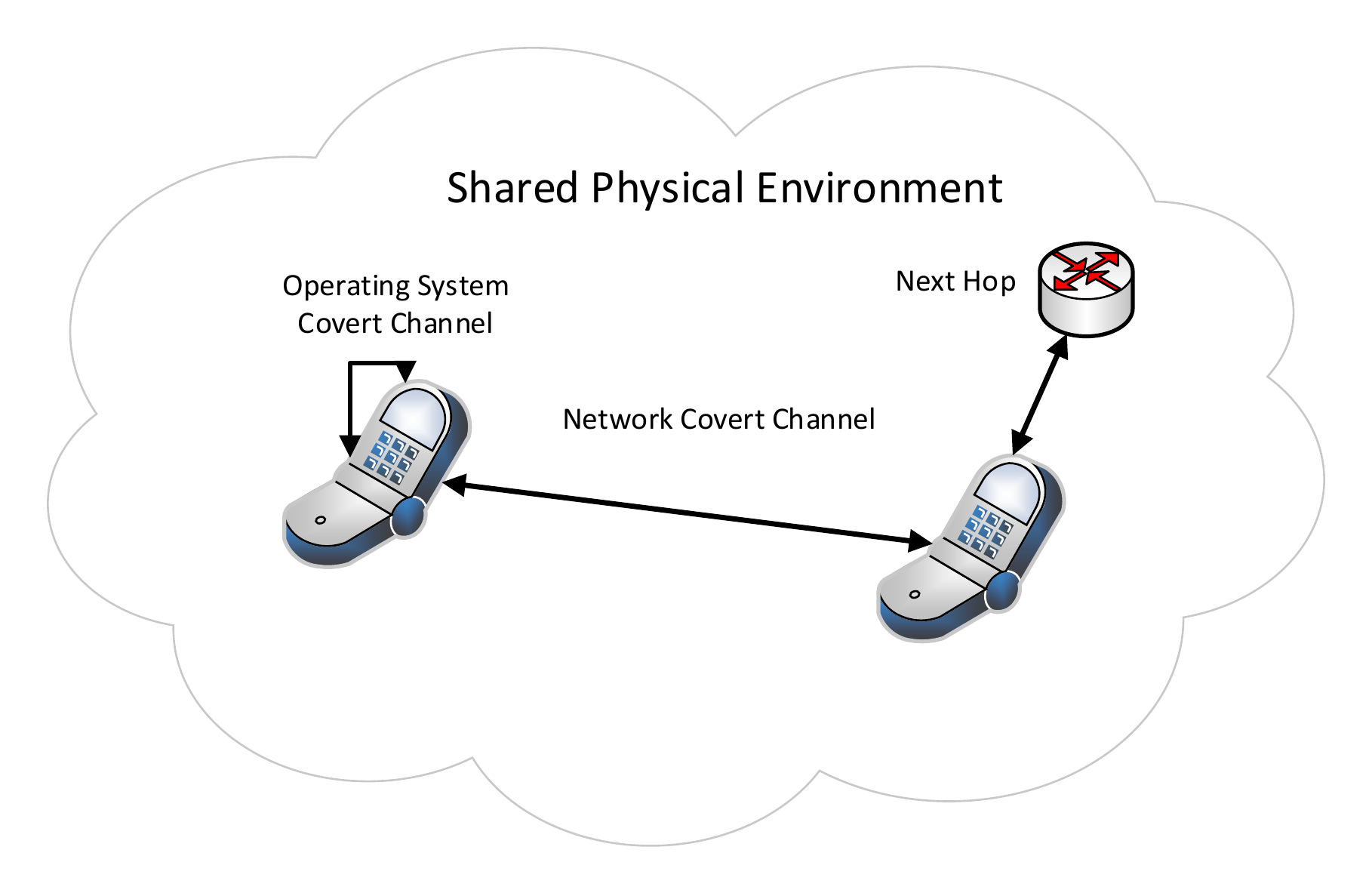}
\caption{Physical Attack Patterns}\label{physical_attack}
\end{figure}

The covert physical channel is used to illegitimately exchange messages between application partitions of a\hfill\newline component-based operating system.
By reutilizing devices that have not been designed for communication at all, the security policy of a pure isolation hypervisor might be circumvented.
In order to establish a covert physical channel, both $p_i$ and $p_j$ must be able to access a device that either transmits physical signals into the computer's environment or receives signals from the computer's physical environment.
Our implementation of a covert physical channel is built upon acoustical signal modulation (utilizing speakers and microphones) but other types of physical signals might also be utilized.
One challenge is to design a covert physical channel with stealthiness in mind in order to prevent early detection.
For this purpose we utilize inaudible acoustical signals (e.g. ultrasound).
Two different physical attack patterns are shown in Fig.~\ref{physical_attack}.

First, it is possible to break the domain separation of a pure isolation hypervisor with a covert physical channel built upon ultrasonic signal modulation.
More specific implementation details on this type of covert physical channel with a bandwidth of up to 600 bit/s and the associated countermeasures will be provided in a future paper.

Secondly, Hanspach and Goetz~\cite{hanspach.jocm} describe an approach where a component-based operating system participates in a covert acoustical mesh network of infected nodes and where the communication range can be extended by transmissions over multiple hops.

\begin{figure*}[!ht]
\includegraphics[width=1.0\textwidth]{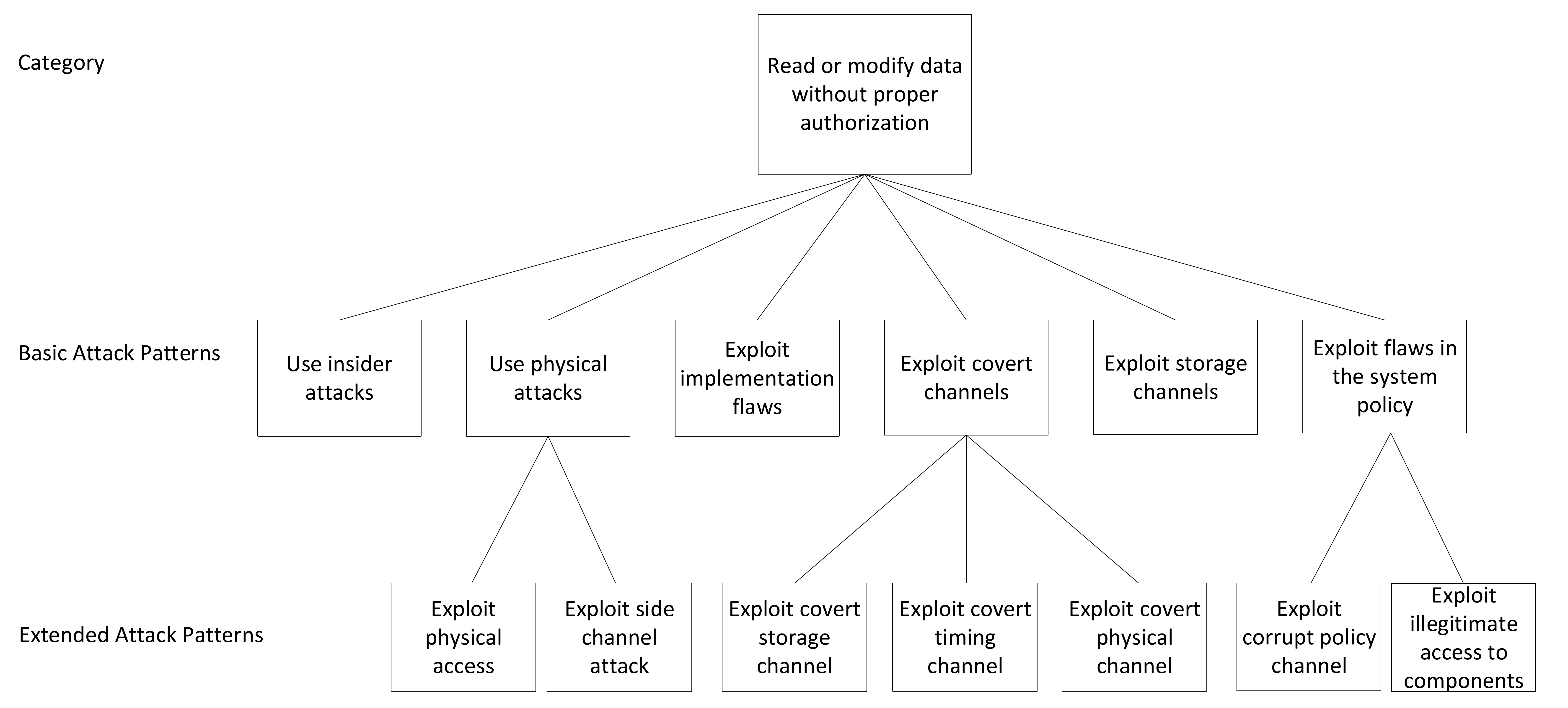}
\caption{Taxonomy of attack patterns in component-based operating systems}\label{attack_tree}
\end{figure*}

Based on the attack patterns in Tab.~\ref{ext_patt2}, we present a taxonomy of attack patterns on information flows in component-based operating systems (modeled after an attack tree \cite{schneier:at}), visualizing the different measures an attacker could implement in order to break the system (Fig.~\ref{attack_tree}).
Each leaf of the tree represents a single attack pattern. 
An attacker seeking to compromise data confidentiality or integrity could be an insider or he could exploit physical means of access, directly accessing the hardware or gathering information from side channels.
Implementation flaws in the hardware or software can affect the operating system's functions, leading to information leaks and violations of the MAC policy of separation in a pure isolation hypervisor.
Moreover, the attacker could create a covert channel, accessing information from partitions, without having a proper clearance for the associated classification level.
Finally, one could attack the operating system by exploiting storage channels, by establishing a policy based channel or by exploiting a flawed system policy to illegitimately get access to system components.

Having defined the possible attack patterns in our scenario, we will now discuss preliminary studies.

\section{Related Work}\label{sec_ap_rl}

Lampson~\cite{Lampson:1973:NCP:362375.362389} made a basic distinction between legitimate channels, storage channels and covert channels in operating systems, which is considerably extended in our work.

Insider attacks in access control systems and countermeasures are described by Baracaldo and Joshi~\cite{Baracaldo:2013:BAU:2462410.2462411}, Liu et al.~\cite{Liu:2008:DSD:1413140.1413159}, and by Yu and Chiueh~\cite{Yu:2004:DFS:1029146.1029154}.
Physical access to a computing system could be exploited to read out unencrypted data after power-down and encrypted data while the system is running.
As Halderman et al.~\cite{Halderman:2009:LWR:1506409.1506429} pointed out, data could even be restored from DRAM for a short time after power-down.
Side channels are presented by van Eck~\cite{vanEck:1985:ERV:7307.7308}, who describes the threat of electromagnetic emanations, by Frankland~\cite{frankland}, who describes optical, acoustical and electromagnetic side channels, and by LeMay, Tan, Shamir and Tromer~\cite{lemay, cryptana}, who study acoustical emanations from computer internal devices (e.g. supply capacitors), while software cache-based side channels are described by~Kong et al.~\cite{10.1109/TC.2012.78}.
Implementation flaws could be used to exploit the computing system by circumventing the MAC policy.
To counter implementation bugs, code-size reduction and formal verification might be applied.
For the kernel level, a formal verified micro kernel has been presented by Klein et al.~\cite{Klein:2009:SFV:1629575.1629596}, while the MILS architectural approach (see Boettcher et al.~\cite{4702758} and Robinson et al.~\cite{Robinson:2007:IMC:1314466.1314474}) aims at providing a trustworthy middleware between the kernel and application partitions, consisting of components that are small enough for evaluation and formal verification.

Lampson \cite{Lampson:1973:NCP:362375.362389} introduces covert channels in 1973, which are defined as communication channels ``not intended for information transfer at all''.
Both covert storage channels and covert timing channels have been described by the National Computer Security Center~\cite{rainbow} in 1993.
Wray~\cite{wray91} presents a study on covert channel terminology where he questions the differentiation between covert storage channels and covert timing channels and concludes that ``storage nature and timing nature are attributes of the channel, and a given channel may possess either or both.''
A covert channel of this combined type could be identified as both a covert storage channel and a covert timing channel in our taxonomy.
In a different approach, Zhai et al.~\cite{Zhai:2009:AIC:1655925.1656006} present a study on automatic identification of covert channels in Linux. 
Kemmerer~\cite{Kemmerer:1983:SRM:357369.357374} introduced the shared resource matrix that describes a methodology to systematically identify covert storage channels in a computing system.
In contrast to the work of Kemmerer, our methodology is not targeted at covert channel identification, but aims to provide a complete information flow analysis in the model of a component-based operating system where different types of attack patterns are identified and described.

Storage channels are also defined by Lampson \cite{Lampson:1973:NCP:362375.362389}, who notes that data is ``written by the service and read by an unconfined program, either shortly after it is written or at some later time'', describing a storage channel between a confined and an unconfined program.
Corrupt policy channels that are present due to a misconfigured access control policy have been addressed by Agreiter~\cite{Agreiter:2008:MCO:1370175.1370221}, proposing automatic generation of an operating system's MAC policy, and Zhai et al.~\cite{Zhai:2009:DIT:1655925.1656007}, presenting a new method for automatic analysis of an operating system's MAC policy.
Finally, exploiting illegitimate access to components would be a special case of a flawed system policy, where a policy circumvents an operating system guard $g_k$ that protects access to a component $c_k$, managing a shared resource, and allows $p_i$ and $p_i$ to exchange messages via $c_k$.
The use of operating system guards to handle shared resources to protect applications from untrusted components has been described by Payne et al.~\cite{Payne:2007:LAS:1278901.1278905}, Robinson et al.~\cite{Robinson:2007:HAM:1228291.1228303,Robinson:2007:IMC:1314466.1314474}, Heckman et al.~\cite{heckman}, and Hanspach and Keller~\cite{hanspach.passat}.

\section{Conclusions}\label{sec_ap6}

We systematically studied and described information flows and patterns to attack these information flows the model of a component-based operating system.
For our basic scenario of component-based operating systems, we analyzed every possible illegitimate information flow, in order to give operating system designers the tools to detect and possibly prevent these illegitimate information flows.
We found very different patterns, ranging from an insider attack to preventing different types of channels and to exploiting implementation flaws, which are mapped to specific information flows.
Any designer of a component-based operating system can use these patterns, to propose new subtypes of attack patterns and develop countermeasures against the associated illegitimate information flows.

The presented algebra might be used in future work to describe and analyze more extended scenarios for information flows in high-assurance setups.
For instance, it might be a good idea to add a variable $E$ for the physical environment of the computing system in order to describe an information flow through the physical environment as present in the covert physical channel and in side channels.
While we analyzed a scenario, where the high-assurance application partition $p_i$ is not connected to a network, a variable $N=\{n_0,\ldots,n_{x-1}\}$ for the network and connected nodes in the networks might be introduced in future work, although complex network-based attack patterns involving $\geq$ 3 hosts might be more appropriately analyzed with a simplified model of the operating system stack.
For network-based attacks, similar attack patterns (e.g. covert channels) could be utilized as described by Wendzel~\cite{wendzeldiss} and many more authors.
As further actors in the scenario, different users of the computing system $U=\{u_0,\ldots,u_{y-1}\}$ might be introduced.
Finally, $H$ might be differentiated to generate a more precise description of attack patterns on information flows between peripheral devices (such as a DMA attack by a malicious network interface card as presented by Duflot et al.~\cite{ssi}).
As we target a mandatory access control system that specifically handles information flows between partitions in a component-based operating system, objects inside a partition (e.g. sockets and files within a VM) are not target of the analysis.

As a general attack pattern against domain separation in a component-based operating system, covert physical channels are introduced, which can be used to physically exchange information between isolated partitions, using light, acoustic wave propagation or any other type of physical signal, which is not already established as a communication channel in the computing system.

In summary, the set of attack patterns presented can be used for a manual for the design and evaluation of a\hfill\newline component-based operating system by addressing each of the identified attack patterns in the security design and evaluation where the operating system model (or a similar model) is applicable.\vfill\eject

\section{Acknowledgements}
We would like to thank our shepherd Rance DeLong for helpful comments toward the final version of this paper.

\bibliographystyle{abbrv}
\bibliography{attack_patterns}

\begin{thebibliography}{10}

\bibitem{Agreiter:2008:MCO:1370175.1370221}
B.~Agreiter.
\newblock Model-driven configuration of os-level mandatory access control:
  research abstract.
\newblock In {\em Companion of the 30th international conference on Software
  engineering}, ICSE Companion '08, pages 995--998, New York, NY, USA, 2008.
  ACM.

\bibitem{anderson:rm}
J.~P. Anderson.
\newblock Computer {Security} {Technology} {Planning} {Study}.
\newblock {\em Technical Report ESD-TR-73-51, Volume II. Electronic Systems
  Divison, AFSC}, Oct. 1972.

\bibitem{Backes:2010:ASA:1929820.1929847}
M.~Backes, M.~D\"{u}rmuth, S.~Gerling, M.~Pinkal, and C.~Sporleder.
\newblock Acoustic side-channel attacks on printers.
\newblock In {\em Proceedings of the 19th USENIX conference on Security},
  USENIX Security '10, pages 20--20, Berkeley, CA, USA, 2010. USENIX
  Association.

\bibitem{Baracaldo:2013:BAU:2462410.2462411}
N.~Baracaldo and J.~Joshi.
\newblock Beyond accountability: using obligations to reduce risk exposure and
  deter insider attacks.
\newblock In {\em Proceedings of the 18th ACM symposium on Access control
  models and technologies}, SACMAT '13, pages 213--224, New York, NY, USA,
  2013. ACM.

\bibitem{4702758}
C.~Boettcher, R.~DeLong, J.~Rushby, and W.~Sifre.
\newblock {The MILS component integration approach to secure information
  sharing}.
\newblock In {\em Digital Avionics Systems Conference, 2008. DASC 2008.
  IEEE/AIAA 27th}, pages 1.C.2--1--1.C.2--14, 2008.

\bibitem{ssi}
L.~Duflot, Y.-A. Perez, V.~Guillaume, and O.~Levillain.
\newblock Can you still trust your network card?
\newblock Agence nationale de la s\`{e}curit\`{e} des syst\'{e}mes
  d`information, Mar. 2010.

\bibitem{Duermuth09}
M.~D{\"u}rmuth.
\newblock {\em Novel classes of side channels and covert channels}.
\newblock PhD thesis, Saarland University, Saarbr{\"u}cken, Germany, 2009.

\bibitem{frankland}
R.~Frankland.
\newblock Side {C}hannels, {C}ompromising {E}manations and {S}urveillance:
  {C}urrent and future technologies.
\newblock Technical Report RHUL-MA-2011-07, Department of Mathematics, Royal
  Holloway, University of London, Egham, Surrey TW20 0EX, England, Mar. 2011.

\bibitem{goguen}
J.~A. Goguen and J.~Meseguer.
\newblock Security policies and security models.
\newblock In {\em Proceedings of the 1982 IEEE Computer Society Symposium on
  Research in Security and Privacy}, pages 11--20. IEEE, Apr. 1982.

\bibitem{Halderman:2009:LWR:1506409.1506429}
J.~A. Halderman, S.~D. Schoen, N.~Heninger, W.~Clarkson, W.~Paul, J.~A.
  Calandrino, A.~J. Feldman, J.~Appelbaum, and E.~W. Felten.
\newblock Lest we remember: cold-boot attacks on encryption keys.
\newblock {\em Commun. ACM}, 52(5):91--98, May 2009.

\bibitem{Halevi:2012:CLK:2414456.2414509}
T.~Halevi and N.~Saxena.
\newblock A closer look at keyboard acoustic emanations: random passwords,
  typing styles and decoding techniques.
\newblock In {\em Proceedings of the 7th ACM Symposium on Information, Computer
  and Communications Security}, ASIACCS '12, pages 89--90, New York, NY, USA,
  2012. ACM.

\bibitem{hanspach.jocm}
M.~Hanspach and M.~Goetz.
\newblock {On Covert Acoustical Mesh Networks in Air}.
\newblock {\em Journal of Communications (in press)}, 2013.

\bibitem{hanspach.passat}
M.~Hanspach and J.~Keller.
\newblock {In Guards we trust: Security and Privacy in Operating Systems
  revisited}.
\newblock In {\em Proceedings of the 5th ASE/IEEE International Conference on
  Information Privacy, Security, Risk and Trust}, PASSAT '13, Washington, DC,
  USA, Sept. 2013. IEEE.

\bibitem{Hasan:2013:SCH:2484313.2484373}
R.~Hasan, N.~Saxena, T.~Halevi, S.~Zawoad, and D.~Rinehart.
\newblock Sensing-enabled channels for hard-to-detect command and control of
  mobile devices.
\newblock In {\em Proceedings of the 8th ACM SIGSAC symposium on Information,
  computer and communications security}, ASIA CCS '13, pages 469--480, New
  York, NY, USA, 2013. ACM.

\bibitem{heckman}
M.~R. Heckman, R.~R. Schell, and E.~E. Reed.
\newblock {Towards Formal Evaluation of a High-Assurance Guard}.
\newblock In {\em Proceedings 6th Layered Assurance Workshop}, pages 25--32.
  \url{http://www.acsac.org/2012/workshops/law/2012-law-proceedings.pdf}, Dec.
  2012.

\bibitem{Jaeger:1998:SAC:319195.319229}
T.~Jaeger, J.~Liedtke, V.~Panteleenko, Y.~Park, and N.~Islam.
\newblock Security architecture for component-based operating systems.
\newblock In {\em Proceedings of the 8th ACM SIGOPS European workshop on
  Support for composing distributed applications}, EW 8, pages 222--228, New
  York, NY, USA, 1998. ACM.

\bibitem{kargerwray}
P.~Karger and J.~Wray.
\newblock Storage channels in disk arm optimization.
\newblock In {\em Proceedings of the 1991 IEEE Computer Society Symposium on
  Research in Security and Privacy}, pages 52--61, 1991.

\bibitem{Karger05multi-levelsecurity}
P.~A. Karger.
\newblock Multi-{Level} {Security} {Requirements} for {Hypervisors}.
\newblock In {\em Proceedings of the 21st Annual Computer Security Applications
  Conference (December 05 - 09, 2005). ACSAC. IEEE Computer Society}, pages
  5--9, 2005.

\bibitem{Kemmerer:1983:SRM:357369.357374}
R.~A. Kemmerer.
\newblock Shared resource matrix methodology: an approach to identifying
  storage and timing channels.
\newblock {\em ACM Trans. Comput. Syst.}, 1(3):256--277, Aug. 1983.

\bibitem{Klein:2009:SFV:1629575.1629596}
G.~Klein, K.~Elphinstone, G.~Heiser, J.~Andronick, D.~Cock, P.~Derrin,
  D.~Elkaduwe, K.~Engelhardt, R.~Kolanski, M.~Norrish, T.~Sewell, H.~Tuch, and
  S.~Winwood.
\newblock {seL4: formal verification of an OS kernel}.
\newblock In {\em Proceedings of the ACM SIGOPS 22nd symposium on Operating
  systems principles}, SOSP '09, pages 207--220, New York, NY, USA, 2009. ACM.

\bibitem{10.1109/TC.2012.78}
J.~Kong, O.~Aciicmez, J.-P. Seifert, and H.~Zhou.
\newblock {Architecting against Software Cache-Based Side-Channel Attacks}.
\newblock {\em IEEE Transactions on Computers}, 62(7):1276--1288, 2013.

\bibitem{Lampson:1973:NCP:362375.362389}
B.~W. Lampson.
\newblock A note on the confinement problem.
\newblock {\em Commun. ACM}, 16(10):613--615, Oct. 1973.

\bibitem{lemay}
M.~D. LeMay and J.~Tan.
\newblock Acoustic {S}urveillance of {P}hysically {U}nmodified {PC}s.
\newblock In {\em Proceedings of the 2006 International Conference on Security
  \& Management}, June 2006.

\bibitem{Liu:2008:DSD:1413140.1413159}
Y.~Liu, C.~Corbett, K.~Chiang, R.~Archibald, B.~Mukherjee, and D.~Ghosal.
\newblock Detecting sensitive data exfiltration by an insider attack.
\newblock In {\em Proceedings of the 4th annual workshop on Cyber security and
  information intelligence research: developing strategies to meet the cyber
  security and information intelligence challenges ahead}, CSIIRW '08, pages
  16:1--16:3, New York, NY, USA, 2008. ACM.

\bibitem{Loughry:2002:ILO:545186.545189}
J.~Loughry and D.~A. Umphress.
\newblock Information leakage from optical emanations.
\newblock {\em ACM Trans. Inf. Syst. Secur.}, 5(3):262--289, Aug. 2002.

\bibitem{Madhavapeddy:2005:ANF:1083818.1083942}
A.~Madhavapeddy, D.~Scott, A.~Tse, and R.~Sharp.
\newblock {Audio Networking: The Forgotten Wireless Technology}.
\newblock {\em IEEE Pervasive Computing}, 4(3):55--60, July 2005.

\bibitem{Murdoch:2006:HRH:1180405.1180410}
S.~J. Murdoch.
\newblock Hot or not: revealing hidden services by their clock skew.
\newblock In {\em Proceedings of the 13th ACM conference on Computer and
  communications security}, CCS '06, pages 27--36, New York, NY, USA, 2006.
  ACM.

\bibitem{rainbow}
{National Computer Security Center}.
\newblock A guide to understanding covert channel analysis of trusted systems.
\newblock \url{http://www.fas.org/irp/nsa/rainbow/tg030.htm}, Nov. 1993.

\bibitem{Payne:2007:LAS:1278901.1278905}
B.~D. Payne, R.~Sailer, R.~C\'{a}ceres, R.~Perez, and W.~Lee.
\newblock A layered approach to simplified access control in virtualized
  systems.
\newblock {\em SIGOPS Oper. Syst. Rev.}, 41(4):12--19, July 2007.

\bibitem{Raguram:2011:IAR:2046707.2046769}
R.~Raguram, A.~M. White, D.~Goswami, F.~Monrose, and J.-M. Frahm.
\newblock {iSpy: automatic reconstruction of typed input from compromising
  reflections}.
\newblock In {\em Proceedings of the 18th ACM conference on Computer and
  communications security}, CCS '11, pages 527--536, New York, NY, USA, 2011.
  ACM.

\bibitem{Robinson:2007:HAM:1228291.1228303}
J.~C. Robinson and J.~Alves-Foss.
\newblock A high assurance {MLS} file server.
\newblock {\em SIGOPS Oper. Syst. Rev.}, 41(1):45--53, Jan. 2007.

\bibitem{Robinson:2007:IMC:1314466.1314474}
J.~C. Robinson, W.~S. Harrison, N.~Hanebutte, P.~Oman, and J.~Alves-Foss.
\newblock Implementing middleware for content filtering and information flow
  control.
\newblock In {\em Proceedings of the 2007 ACM workshop on Computer security
  architecture}, CSAW '07, pages 47--53, New York, NY, USA, 2007. ACM.

\bibitem{Rushby:1981:DVS:800216.806586}
J.~M. Rushby.
\newblock Design and verification of secure systems.
\newblock In {\em Proceedings of the eighth ACM symposium on Operating systems
  principles}, SOSP '81, pages 12--21, New York, NY, USA, 1981. ACM.

\bibitem{schneier:at}
B.~Schneier.
\newblock Attack {Trees}.
\newblock {\em Dr. Dobb's Journal}, Dec. 1999.

\bibitem{cryptana}
A.~Shamir and E.~Tromer.
\newblock {Acoustic cryptanalysis: On nosy people and noisy machines. The
  Blavatnik School of Computer Science, Tel Aviv University}.
\newblock \url{http://cs.tau.ac.il/~tromer/acoustic/}.

\bibitem{vanEck:1985:ERV:7307.7308}
W.~van Eck.
\newblock Electromagnetic radiation from video display units: an eavesdropping
  risk?
\newblock {\em Comput. Secur.}, 4(4):269--286, Dec. 1985.

\bibitem{watson_mac}
R.~N.~M. Watson.
\newblock {A Decade of OS Access-control Extensibility}.
\newblock {\em Queue}, 11(1):20:20--20:41, Jan. 2013.

\bibitem{wendzeldiss}
S.~Wendzel.
\newblock {\em {Novel Approaches for Network Covert Storage Channels}}.
\newblock PhD thesis, FernUniversit{\"a}t in Hagen, Faculty for Mathematics and
  Computer Science, Hagen, Germany, 2013.

\bibitem{wray91}
J.~Wray.
\newblock An analysis of covert timing channels.
\newblock In {\em Proceedings of the 1991 IEEE Computer Society Symposium on
  Research in Security and Privacy}, pages 2--7, May 1991.

\bibitem{Yu:2004:DFS:1029146.1029154}
Y.~Yu and T.-c. Chiueh.
\newblock Display-only file server: a solution against information theft due to
  insider attack.
\newblock In {\em Proceedings of the 4th ACM workshop on Digital rights
  management}, DRM '04, pages 31--39, New York, NY, USA, 2004. ACM.

\bibitem{Zhai:2009:DIT:1655925.1656007}
G.~Zhai, W.~Ma, M.~Tian, N.~Yang, C.~Liu, and H.~Yang.
\newblock {Design and implementation of a tool for analyzing SELinux secure
  policy}.
\newblock In {\em Proceedings of the 2nd International Conference on
  Interaction Sciences: Information Technology, Culture and Human}, ICIS '09,
  pages 446--451, New York, NY, USA, 2009. ACM.

\bibitem{Zhai:2009:AIC:1655925.1656006}
G.~Zhai, Y.~Zhang, C.~Liu, N.~Yang, M.~Tian, and H.~Yang.
\newblock Automatic identification of covert channels inside linux kernel based
  on source codes.
\newblock In {\em Proceedings of the 2nd International Conference on
  Interaction Sciences: Information Technology, Culture and Human}, ICIS '09,
  pages 440--445, New York, NY, USA, 2009. ACM.

\end{thebibliography}

%
\end{document}